# Solenoid-free current drive via ECRH in EXL-50 spherical torus plasmas

Yuejiang Shi*, Bing Liu*, Shaodong Song, Yunyang Song, Xianming Song, Bowei Tong, Shikui Cheng, Wenjun Liu, Mingyuan Wang, Tiantian Sun, Dong Guo, Songjian Li, Yingying Li, Bin Chen, Xiang Gu, Jianqing Cai, Di Luo, Debabrata Banerjee, Xin Zhao, Yuanming Yang, Wenwu Luo, Peihai Zhou, Yu Wang, Akio Ishida, Takashi Maekawa, Minsheng Liu, Baoshan Yuan, Y-K Martin Peng* and the EXL-50 team

Hebei Key Laboratory of Compact Fusion, Langfang 065001, China

Enn Science and Technology Development Co., Ltd, Langfang 065001, China

*E-mail of corresponding author: shiyuejiang@enn.cn/yjshi@ipp.ac.cn  liubingw@enn.cn  pengyuankai@enn.cn

**Abstract**

As a new spherical tokamak (ST) designed to simplify engineering requirements of a possible future fusion power source, the EXL-50 experiment features a low aspect ratio (A) vacuum vessel (VV), encircling a central post assembly containing the toroidal field coil conductors without a central solenoid. Multiple electron cyclotron resonance heating (ECRH) resonances are located within the VV to improve current drive effectiveness. Copious energetic electrons are produced and measured with hard X-ray detectors, carry the bulk of the plasma current ranging from 50kA to 150kA, which is maintained for more than 1s duration. It is observed that over one Ampere current can be maintained per Watt of ECRH power issued from the 28-GHz gyrotrons. The plasma current reaches $I_p$>80kA for high density (>$5\times10^{18}$m$^{-2}$) discharge with 150kW ECHR heating. An analysis was carried out combining reconstructed multi-fluid equilibrium, guiding-center orbits of energetic electrons, and resonant heating mechanisms. It is verified that in EXL-50 a broadly distributed current of energetic electrons creates smaller closed magnetic-flux surfaces of low aspect ratio that in turn confine the thermal plasma electrons and ions and participate in maintaining the equilibrium force-balance.

1. Introduction

Great progress has been achieved in magnetic confinement fusion research based on the tokamak since 50 years ago when the first stable high temperature plasma was observed in the T-3 tokamak [1,2]. The tokamak has been the most investigated and furthest advanced configuration among the magnetic confinement fusion systems. More recently, the spherical tokamak (ST) concept of aspect ratios around 1.5 [3, 4] has been experimentally (START [5], NSTX [6], MAST [7], and Globus-M [8]) tested to realize a substantially higher plasma beta compared to the tokamak of aspect ratios around 3, and is an attractive candidate for realizing a relatively compact fusion reactor. In this article, the special torus (ST) indicates the spherical tokamak.

The tokamak plasma current is required to insure a high plasma confinement capability to restrain transport losses from the core to the edge. The start-up and ramp-up of this current have been commonly driven by a toroidal electric field induced by current changes in a centre solenoid (CS) magnet. This however causes engineering difficulties for the ST due to the limited space available within a narrow centre column. Furthermore, a CS magnet is capable of sustaining the plasma current over limited time period, which is to be augmented by non-inductive methods in a future fusion reactor. To develop a solenoid-free current drive capability therefore has been an important research endeavour for the STs. On the positive side, removing the CS allows additional space to increase the toroidal field (TF), further improving compactness and economy.

The original physics concept and principle of the ENN Spherical Torus with a major radius 58 cm (EXL-50) in Energy iNNovation (ENN) Science and Technology Development Co. were recently proposed by Peng [9]. One of the key EXL-50 experimental goals is to test the effectiveness of electron cyclotron resonance heating (ECRH) and current drive in the absence of an CS magnet. CS-free ECRH and current drive have been tested in several earlier ST devices (CDX-U [10], LATE [11-15], TST-2 [16-17], MAST [18-19], and QUEST [20-26]). A toroidal current of 1.05 kA was generated using about 8 kW of ECRH power on CDX-U [9], proving the possibility of current start-up by ECRH alone. Later, a 7kA plasma current was generated by about 30kW ECRH in LATE [11-12]. A current flattop with closed flux surface (CFS) plasma was sustained for 60ms in LATE, proving the potential for steady-state ECRH and current drive of the ST plasmas. In MAST, a plasma current of 73kA was produced by 60kW ECRH power with the help of the unique grooved mirror-polarizer installed on the central rod [18]. In QUEST, a plasma current of 90kA was obtained with about 200kW ECRH power through combined first and second harmonic resonances [24].

In this paper, we present the latest ECRH experimental results from EXL-50. Not only are the operational parameters of CS-free current drive by ECRH significantly expanded, but also observed are some remarkable plasma behaviours. Discharges with plasma currents substantially above 100kA are routinely obtained in EXL-50, with the current flat-top sustained for up to or beyond 2 seconds. Data of current drive efficiency higher than 1A current per Watt of ECRH power issued from the gyrotrons, averaged over hundreds of discharges, have been accumulated. Plasma currents as high as 80-100 kA have been achieved at line-densities over $0.5 \times 10^{19} m^{-2}$.

This paper is organized as follows: an introduction to the experiment setup in EXL-50 is given in section 2. The high efficiency current drive experimental results are described in section 3. Section 4 gives the discussion of energetic electrons and current drive mechanisms. High density current drive experiments are presented in section 5. Conclusions and future plans are summarized in section 6.

2. Experimental setup

The EXL-50 device is a medium-sized ST with a cylindrical vacuum vessel (see, Fig. 1). An important characteristic of EXL-50 is that it does not have a central solenoid. Six poloidal field (PF) coils are located outside the vacuum vessel and the TF coil conductors. Inner limiters on the center column and outer limiters on the vessel wall have leading edges at 0.186m and 1.512m in major radius, respectively. The design of the large space of EXL-50's vacuum vessel is mainly for the confinement and accommodation of energetic electrons whose spatial distribution area is larger than that of thermal plasmas. Two microwave frequencies have been utilized so far, 28-GHz from high power gyrotrons for higher toroidal field discharges and 2.45-GHz from low power magnetrons for lower toroidal field discharges and wall cleaning. Fig.1 shows the poloidal cross section of the EXL-50 device. Two sets of 28GHz gyrotrons (50kW source power for ECRH1 and 400kW for ECRH2) are available to inject power through two outboard ports above the mid-plane. Another 400kW 28GHz gyrotron (ECRH3) and two sets of 2.45GHz magnetrons (30kW source power each) are available to inject power through the mid-plane ports. The toroidal injection angles of the ECRH systems can be adjusted over limited ranges (as shown in Fig.1b). Both the 2.45-GHz and the 28-GHz systems are arranged to inject primarily ordinary-mode (O-mode) wave in recent experiments on the EXL-50. When the electric current of a 12-turn TF coils per turn was set to about 100kA, the fundamental and higher ECR layers (up to five resonances) coexist within the EXL-50 vacuum region (as shown in Fig.1a). The electron density is measured by single-chord tangential microwave interferometer [27]. Two CdTe detectors with energy resolution are applied to observe the forward and backward bremsstrahlung hard x-ray (HX) emission [28].

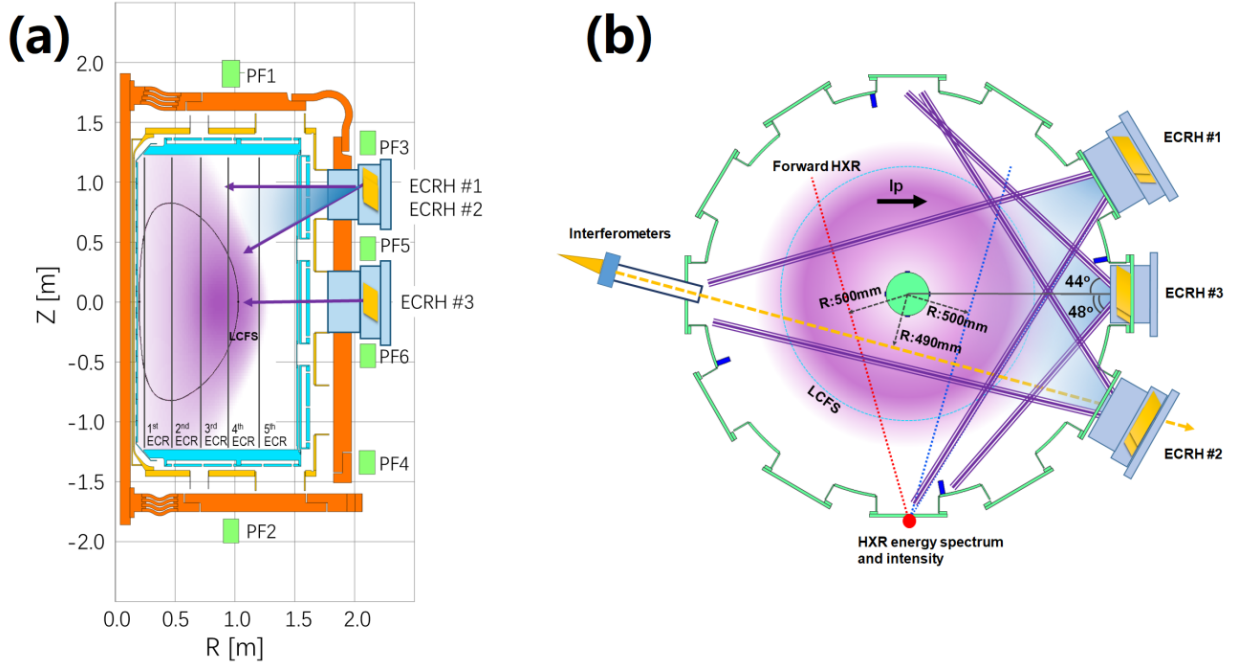

Fig.1 (a) Poloidal cross section of the EXL-50 device. Nominal toroidal field is 1 T at R = 0.24m. (b) Top view of EXL-50. The lines of sight of interferometer and HX diagnostics are indicated in the figure. The ECRH beam is aimed at the center of the machine when the toroidal injection angles is $0^0$. The typical plasma cross section (purple cloud) is also shown in fig.1a and fig.1b.

## 3. High efficiency current drive experimental results in EXL-50

Here, a simplified current drive effectiveness $\eta_{A/W}$ is defined as follows and utilized:

$$\eta_{A/W} = I_P/P_{ECRH}$$

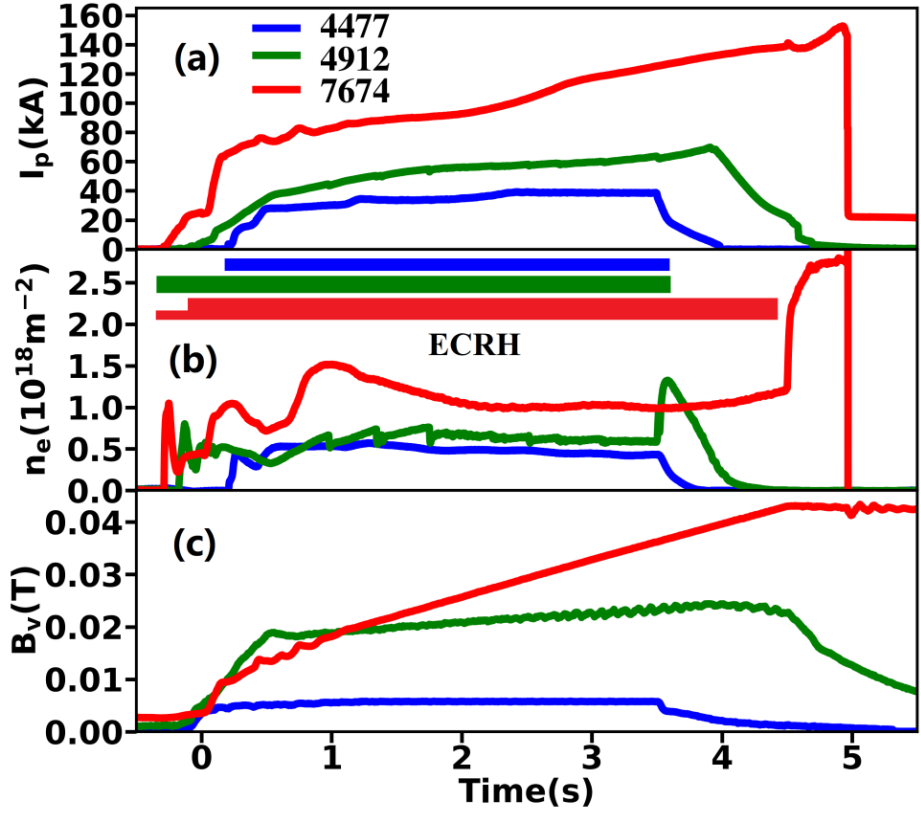

Fig.2 The discharge waveforms for different 28GHz ECRH heating power. (a) plasma current; (b) line integrated density $P_{ECRH}$ was 20kW in shot 4477 and 45kW in shot 4912, respectively. Two gyrotrons are used in shot 7674. One gyrotron injected 20kW from -0.3s to 0s and the other gyrotron injected 115kW from 0s to 4.5s.

where $I_P$ is the plasma current, $P_{ECRH}$ is the ECRH power issued from the gyrotrons. Fig.2 shows the typical discharge waveforms with different 28GHz ECRH heating power in EXL-50. The $\eta_{A/W}$ can reach 2A/W (40kA/20kW) for low ECRH power plasma. The $\eta_{A/W}$ is 1.55A/W (70kA/45kW) for moderate ECRH power plasma and 1.22A/W (140kA/115kW) for further increased ECRH power plasma. The major and minor radius during flattop phase for the thermal plasma inside last closed magnetic surface are 0.59m and 0.41m for shot 4477, 0.44m and 0.26m for shot 4912, 0.52m and 0.34m for shot 7674, respectively. The line averaged density during flattop phase is around $0.26 \times 10^{18} m^{-3}$, $0.61 \times 10^{18} m^{-3}$, and $0.80 \times 10^{18} m^{-3}$ for the three shots in fig.2, respectively. The $P_{ECRH}$ in this paper is the power measured at the matching optical unit (MOU) which is close to the exit power of the gyrotrons. The power delivered from the antenna inside the vacuum vessel is unknown at present due to the lack of monitoring equipment. A directional coupler in the miter bend will be installed to obtain the waveform of the injected power in future experiments. The duration of high current ($I_P$ >100kA) for the higher ECRH power plasma in Fig.2 is longer than 2s. The total pulse length for 28GHz ECR heating plasma is less than 6s limited by temperature rise at the top joints of the TF coils at 100kA current. One notable phenomenon shown in Fig.2 is the density jump when the ECRH power is turned off, indicating possibly a cessation of density pump-out by ECH [29] or confinement transition, which is not addressed in this paper.

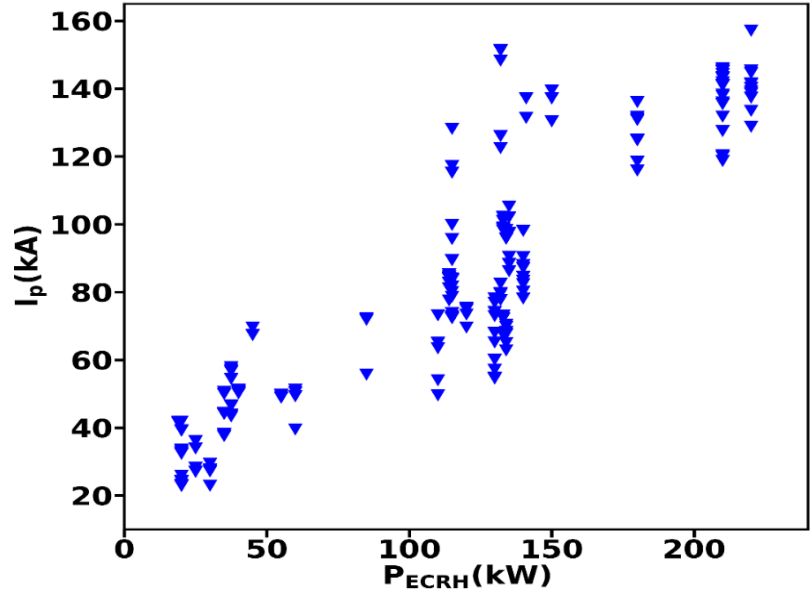

Fig.3 $I_p$ v.s. $P_{ECRH}$ for 200 successful shots in EXL-50

Fig.3 shows the relation of $I_p$ v.s. $P_{ECRH}$ for about 200 successful shots in EXL-50. The general trend in Fig.3 is that the Ip increases with $P_{ECRH}$. On the other hand, it can also be seen in Fig.3 that the $I_p$ varies in quite large range for the same $P_{ECRH}$. The uncertainty of actual injected and absorbed power by plasma may be one reason for the scattering relation between Ip and $P_{ECRH}$ in Fig.3. At the same time, changes of the currents in PF coils and the vertical magnetic field $B_v$ have substantial effects on $\eta_{A/W}$. For the same coil currents, $B_V$ (at R=0.5 in middle plane) contributed from PF5 & 6 is around 6 times as high as that of PF12 and 2 times as high as that from PF3 & 4. So, we have performed the special discharge experiments, as shown in fig.4a. In these discharges, only the current in PF5 & 6 are changed to show the effect of $B_v$ on plasma current while the other parameters are the same in order to show the effect of $B_v$ on plasma currents. . Fig.4a clearly shows that the $I_p$ increases with the current in PF5 & 6 coils for the same $P_{ECRH}$ and density. The force balance between the expansion of the plasma itself and external magnetic compress must be sustained for a stable equilibrium configuration. More statistical information from 200 shots is shown in Fig.4b. It can be found that the $I_p$ increases with the external vertical magnetic field $B_v$ in the appropriate $P_{ECRH}$ range. Fig.4b also demonstrates that both $B_v$ and $P_{ECRH}$ are the essential elements for increasing the plasma current. $B_v$ is not a plasma current driving source, but it will affect the maximum plasma current driven by ECRH. So much potential for raising $\eta_{A/W}$ through optimizing and matching of PF coil current and power of ECRH remains unexplored at present, which will be explored and improved in future experiment in EXL-50. In the first experiment campaign in EXL-50, the main target are to start-up and maintain plasma current. The density is operated in a narrow range ($0.5\sim2\times10^{18}m^{-2}$) for the shots in fig.3. Fig.4c shows the relation between the plasma current and the density in flattop phase for the same shots in fig.3 and fig.4b. It can be seen that too low density is not conducive to the increase of plasma current. The favorite density for high Ip increases with $P_{ECRH}$.

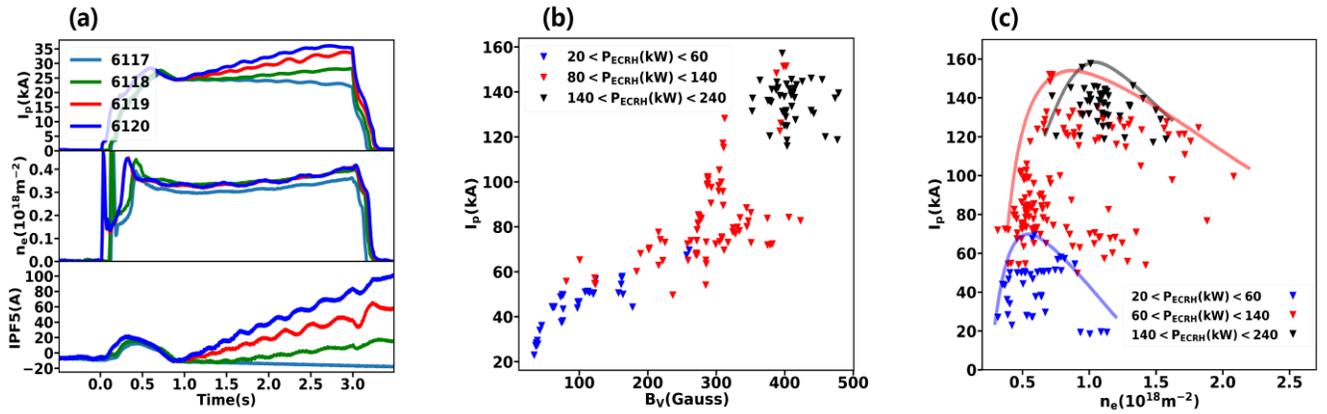

Fig.4 (a) The waveforms of four discharges with similar density. The waveforms from top to bottom are plasma current, line integrated density, and the electrical current in PF5 coil. $P_{ECRH}$ was 25kW for these shots. The current in PF1 and PF3 were kept constant ($I_{PF1}$=100A and $I_{PF3}$=600A) during whole discharge phase. (b) The plasma current in flattop phase versus external $B_v$ at R=0.5m in middle plane. (c) The plasma current versus line integrated density.

## 4. Energetic electrons and current drive mechanisms

The Pfirsch-Schluter (PS) current is a dominant component during the initial start-up phase, and drastically decreases with increasing $B_v$ following the formation of CFS. The boot-strap current drive by the pressure gradient is at present estimated to be less than several percent for these EXL-50 plasmas. The conventional electron cyclotron current drive (ECCD) via Fisch-Boozer mechanism [30] or Ohkawa mechanism [31] can also contribute to the non-inductive current. However, such ECCD effects, being sensitive to the ECRH injection angle, have not been confirmed in EXL-50 experiments. Fig.5 shows the waveforms of two shots with the same $P_{ECRH}$ in EXL-50. Although the toroidal angle for the ECRH antenna was set at $-16^0$ for count-current drive in shot 7448 and $17^0$ for co-current drive in shot 7449, the plasma current remained largely unchanged. The single pass absorption of electron cyclotron wave (ECW) is very weak in the present low temperature EXL-50 plasmas. The angle and mode of ECW are randomized during the multiple wall reflections, so that the conventional ECCD mechanism may contribute a negligible fraction to the total plasma current.

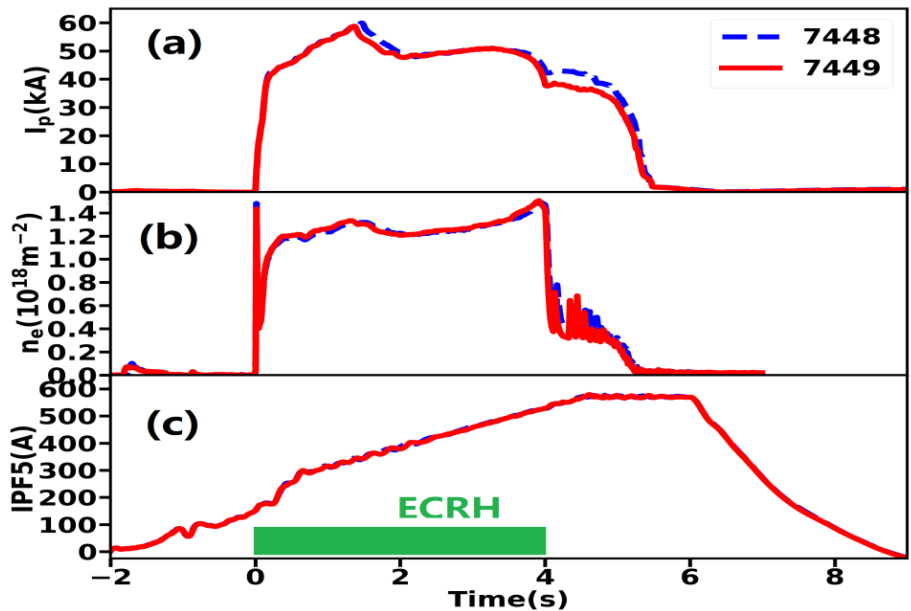

Fig.5 The waveforms of discharges with same density, ECRH power and PF coil current. (a) Plasma current; (b) The current in PF5; (c) line integrated density; The toroidal angle of ECRH antenna is co-current direction in shot 7448 and counter-current direction in shot 7449.

QUEST and LATE experiments have proven that the energetic electrons play a primary role for the CS-free current drive. EXL-50 experimental results confirm that the plasma current is mainly carried by such energetic electrons. The shot shown in Fig.6 is a very stable and well-controlled discharge, showing a nearly stationary plasma current, electron density, as well as a zero loop voltage from 1.5s to 4.5s. During the entire discharge, the plasma current, and hard x-ray intensity and its photon temperature (the average energy of energetic electrons) vary conjointly in magnitude. It is seen that both the number and the energy of energetic electrons contribute directly to the increase of plasma current.

It should be noted that the role of induction in the CS-free ECRH driven current remains unresolved. That is, does the toroidal electric field induced by changes in the PF coils and plasma currents accelerate the already decoupled energetic electrons to even higher energies and carry a significant fraction of plasma current during a discharge? Experiments dedicated to resolving this question were carried out. As indicated in Fig. 6a, the currents in PF3 and PF4 (not indicated) were kept constant during the entire discharge. The currents of PF1&2 and PF5&6 were ramped-up slowly until 1.5s and kept constant through to 6s. The loop voltage oscillated near zero between 1.5s and 2.5s and became zero between 2.5s and 4.5s. The loop voltage, measured at the mid-plane of low field side, is the main source for driving inductive plasma current. The definition of polarity for loop voltage and Ip in EXL50 is reversed to each other. A positive loop voltage means counter-clock wise direction from top view. A positive Ip means clock wise direction from top view. The largest value of loop voltage of +0.17V appears at 0.037s, in the direction opposite to the plasma current. Therefore, the contribution to the inductive plasma current from loop voltage has a negligible or even negative contribution to the total plasma current during start-up and flattop phases.

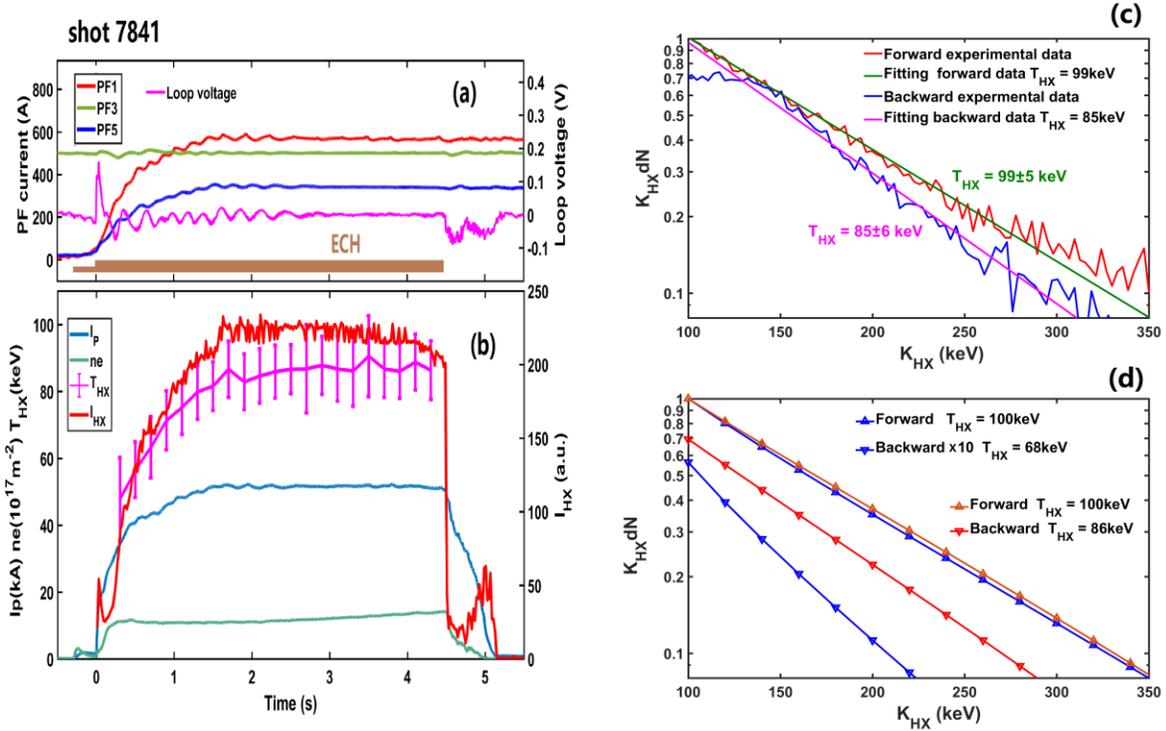

Fig.6 (a) Waveforms of PF currents (PF1, PF3, and PF5), loop voltage and ECRH. (b) Waveforms of plasma current (Ip), density (ne), temperature (THX) and intensity (IHX) of hard x-ray. One gyrotron injected 10kW from -0.3s to 0s and the other gyrotron injected 100kW from 0s to 4.5s. (c) HX spectrum (integrated time: 2s~2.2s) in forward and backward direction during flattop phase. The data are normalized with intensity with the intensity at 100keV of forward HX. (d) Simulated HX spectrum with three-temperature distribution model for energetic electron. Blue line (runaway case) : $T_{//F}=10T_\perp =10T_{//B}$. The backward data is magnified 10 times. The data is normalized with intensity at 100keV of forward HX. Red line: $T_{//F}=T_\perp$ , $T_{//B} =0.75T_{//F}$. The data are normalized with intensity at 100keV of forward HX.

In addition, the velocity distribution of energetic electrons driven by ECRH is different from that of runaway electrons induced by a toroidal electric field. In the former case, the energetic electrons possess similar magnitudes of parallel and perpendicular velocities, while in the latter, the parallel velocities dominate. Fig.6c shows the hard x-ray intensity and energy spectrum in forward (count-current) direction and backword (co-current) direction, indicating relatively moderate differences. The three-temperature Maxwellian distribution model (3T model) [32, 33] can be applied for the anisotropic distribution of runaway electrons or energetic electrons in RF heating plasmas. The detail for the HX spectrum simulation is presented in appendix I. The simulated HX spectrum based on 3T model is shown in Fig.6d. For the runaway distribution case ($T_{//F}=10T_\perp =10T_{//B}$, blue line in Fig.6d), the simulated backward HX intensity at 100keV is around 1/20 that of forward HX intensity. The backward photon temperature of HX is only as 2/3 as that of forward HX in the runaway cases. Compared to the runaway simulation cases, the simulated HX spectrum based on the distribution of equal parallel forward and perpendicular temperature (red line in Fig.6d) is much more approximateto the experimental data in Fig.6c. On the other hand, the energetic electrons can be strongly accelerated in the parallel direction by the relative high loop voltage during current ramp-down after the ECRH is turned off. These experimental observations indicate that the inductive and runaway-like current drive mechanisms are not significant in the CS-free ECRH plasmas.

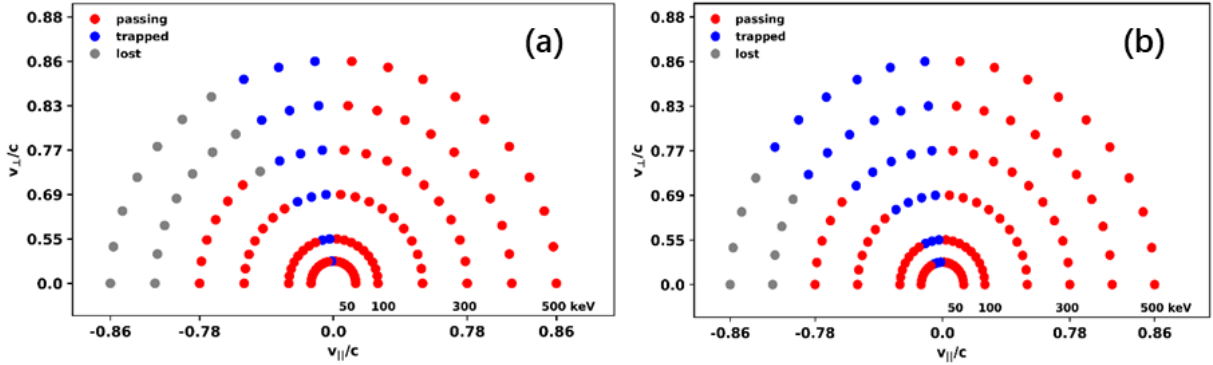

Fig.7 (a) Velocity distribution of energetic electrons which start from 1st harmonica layer at mid-plane. (b) Velocity distribution of energetic electrons which start from magnetic axis. The background magnetic field for the simulation of energetic electrons is obtained from the three-fluid equilibrium model. The typical three-fluid equilibrium structure can be seen in fig.8.

A key question regarding the very high current drive effectiveness ($\eta_{A/W}$ = 1.2 ~2 A/W) in EXL-50 via the energetic electrons may be addressed by analysing the asymmetric region of orbit containment of the energetic electrons. Fig.7 shows the orbit confinement analysis for energetic electrons in velocity space. A three-fluid equilibrium of a 50kA EXL-50 plasma was obtained via the multi-fluid equilibrium model [33] for the computation of the guiding-center orbits [34]. A strongly asymmetric distribution in the parallel direction of the contained orbits in the $v_\parallel$, $v_\perp$, and energy space, is obtained, and shown to be accentuated as the electron energy increases toward the limiting energy of orbit containment. The asymmetric structure of the confined energetic electron orbits as shown in fig.7 is determined by the PF coil currents, not the ECRH injected angle. The population of energetic electrons is mainly related to the density and power of ECRH. As mentioned above, the conventional ECCD mechanisms provide a minor contribution to the total plasma current in EXL-50. Although the ECRH injection angles are quite different in the shots in fig.5, the plasma currents are quite similar because the other main parameters such as the PF setting, density and ECRH power are the same. Further, the design of EXL-50 (as shown in Fig.1a) permits the coexistence of five ECR layers within the vacuum vessel. Considering the effect of relativistic Doppler shift [23], the resonance layers for energetic electrons broaden in major radius direction. Fig.8a shows the radial dependence of the characteristic resonant energies for the fundamental and harmonic ECW for EXL-50. For electrons of energy above 100keV, the width of a resonance upshifts to overlap with the downshifted resonance of the next higher harmonic. It can be seen that the individual resonance widths for the energetic electrons fill the entire space inside EXL-50's vacuum vessel.

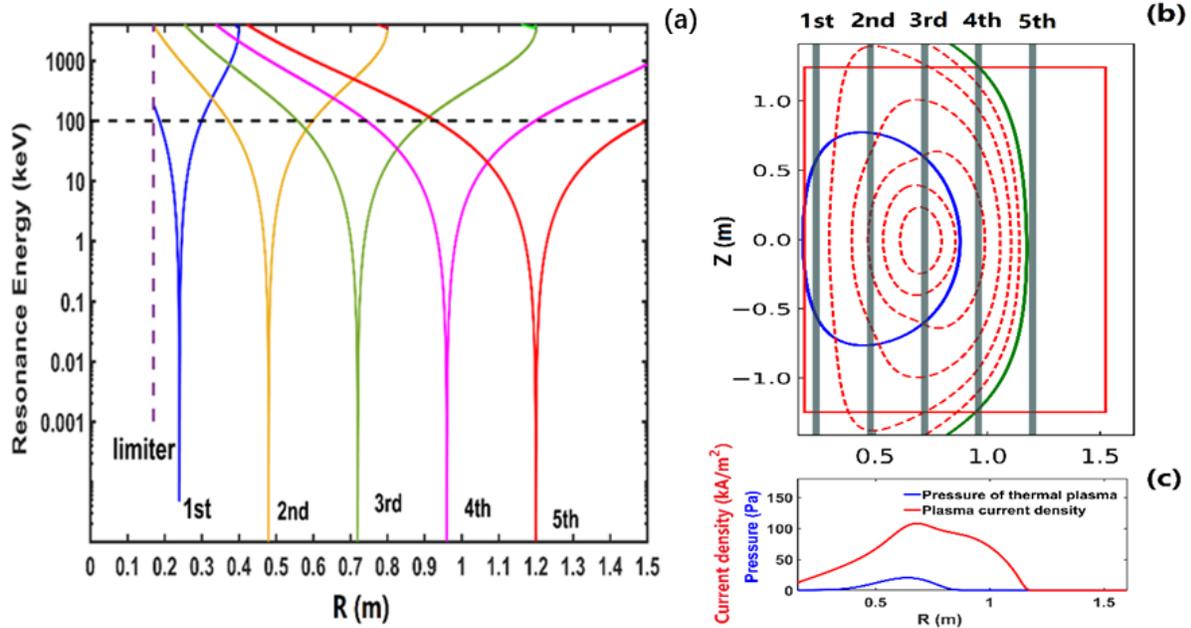

Fig.8 (a) Radial profiles of resonance energy for the harmonic ECW. (b) Reconstructed 2D contour plot of plasma current by the multi-fluid equilibrium model. Blue line represents LCFS. (c) The radial profiles of thermal plasma pressure and current density

The single-pass absorption of ECW is estimated to be relatively low in EXL-50 for the present range of plasma densities and temperatures. The smooth stainless steel vacuum vessel walls and limiters, including those on the center column, assist in ensuring multiple reflecting paths of the injected ECW back to the plasma. Wall-reflection further helps by converting O-mode wave to the X-mode and vice versa, thus taking advantage of the higher efficiency of X-modes by energetic electrons [26]. Another notable feature of EXL-50 plasma is that the cross section of the plasma current carried by the energetic electrons is much bigger than that of CFS during the flattop phase of plasma current. Fig.8b and c show 2D contour plot of plasma current, last close flux surface (LCFS), and the profiles of thermal plasma and current for a 120kA discharge shown in Fig.2, computed via the multi-fluid equilibrium model [30]. A significant fraction of the plasma current (52% in this case) is flowing outside the LCFS. The phenomenon that the energetic electrons play a substantial role in the formation of closed flux surface and carry a dominant fraction of the plasma current that extends over the open field line region in the solenoid-free ECRH sustained ST plasmas has been confirmed in LATE [11,12]. Similar analysis has also been conducted in EXL-50. As a first approximation, a multi-fluid equilibrium model that includes a high-energy electron component in addition to the low-energy electron and ion components was applied to describe the equilibrium characteristics of EXL-50. The simulation results were in good agreement with the available experimental data [33]. Some special experiments have been performed in EXL-50 to indicate the considerable population of energetic electrons outside the LCSF [35]. Moreover, a removable limiter will be installed in future machine upgrade plan. The space for high harmonic ECH resonance layer and gap between LCFS and limiter will be actively controlled to systematically investigate their effects on plasma current in EXL-50.

## 5. Current drive in high density plasmas in EXL-50

High density ECRH discharges are also obtained in EXL-50. The electron Bernstein wave (EBW) can be excited and played key role to heat plasma and drive current for the over-dense (i.e. the density is higher than the cut-off density of electron cyclotron wave ) ECRH plasmas. Such over-dense plasma has been achieved in the low toroidal field (TF=20kA) operation mode in EXL-50. The 2.45Ghz microwave system is applied for the ECRH heating when the current of TF coil reduce to 20kA. There are still two resonance layers coexist for the 2.45GHz ECRH in low TF situation. The major radius for the three resonance layers are 0.55 m and 1.1 m for the $1^{st}$ and $2^{nd}$ harmonic, respectively. The typical discharge waveforms of over-dense plasma with 2.45GH ECRH heating are shown in fig.9a. The line averaged density can be as higher as three times of the ordinary mode (O-mode) cut-off density in the 2.45GHz ECH discharges.

High density plasma is obtained with multi-pulse gas puffing in 28GHz ECRH discharges. The nozzle for gas puffing is located at the middle of the central stack (Z=0 in fig.1a). It can be seen in fig.10 that the density increase rapidly at round 3s. The plasma density keeps high level ($>5\times10^{18}$m$^{-2}$) status from 3s to 4s. At the same time, the plasma current is higher than 80kA during high density phase. The density profile for core plasma is unavailable in EXL-50 at present. The core density maybe exceeds the cut-off density of 28GHz ECW in the high density plasma. It is surmised that the electron Bernstein wave (EBW) has been excited to drive current for such high density discharges.

The general current drive efficiency $\eta_{CD}=n_eRI_p/P_{ECRH}$ for non-inductive current is estimated for the high density ECRH experiments in EXL-50. For the 2.45GHz ECRH high density discharge in fig.9, $\eta_{CD}=n_eRI_p/P_{ECRH} = 0.013\times10^{19}\times0.53\times3kA/20kW$ is around $0.011\times10^{19}$ MA MW m$^{-2}$. The $\eta_{CD}=n_eRI_p/P_{ECRH} = 0.43\times10^{19}\times0.52\times85kA/150kW$ is around $0.13\times10^{19}$ MA MW m$^{-2}$ for the 28GHz ECRH high density plasma fig.10. It looks like that the high TF or ECRH frequency has tremendous benefit to improve the current drive efficiency. Here, we make a simple and bold extrapolation for the ECRH current drive efficiency in STs based on the EXL-50's experimental results. We assume that the $\eta_{CD}$ of ECRH in EXL-50 like STs has strong correlation with the wave frequency $f_{ECRH}$ or toroidal magnetic field $B_T$, i.e.

$$\eta_{CD} \propto f_{ECRH}^{\alpha}, \quad or \quad \eta_{CD} \propto B_T^{\beta}$$

On the other hand, the ECRH frequency should match with the TF. Base on the two sets experimental results on EXL-50, the indexes α and β in above formula are estimated to be around 1.0 and 1.5, repsectively. We also assume that $\eta_{CD}$ can be lineally extrapolated to high $B_T$ and high $f_{ECRH}$ STs from EXL-50 results, i.e.,

$$\eta_{CD} = \eta_{CD}^{EXL-50}(f_{ECRH}/f_{ECRH}^{EXL-50}), \quad or \quad \eta_{CD} = \eta_{CD}^{EXL-50}(B_T/B_T^{EXL-50})^{1.5}$$

For a high $B_T$ (3.5T) and high $f_{ECRH}$ (170GHz) EXL-50 like STs, $\eta_{CD}$ will reach $0.79\times10^{19}$ MA MW m$^{-2}$ based on frequency relation, or $2.4\times10^{19}$ MA MW m$^{-2}$ based on TF relation. For a reactor size 3.5T ST (R=1m), 4~10MW 170GHz ECRH can drive around 5MA plasma current with density around $2\times10^{19}$m$^{-3}$ in start-up phase. We emphasize that the above estimates for ECRH current drive efficiency and power to start-up the plasma current in reactor size ST are very preliminary and superficial, and there maybe some omissions. .

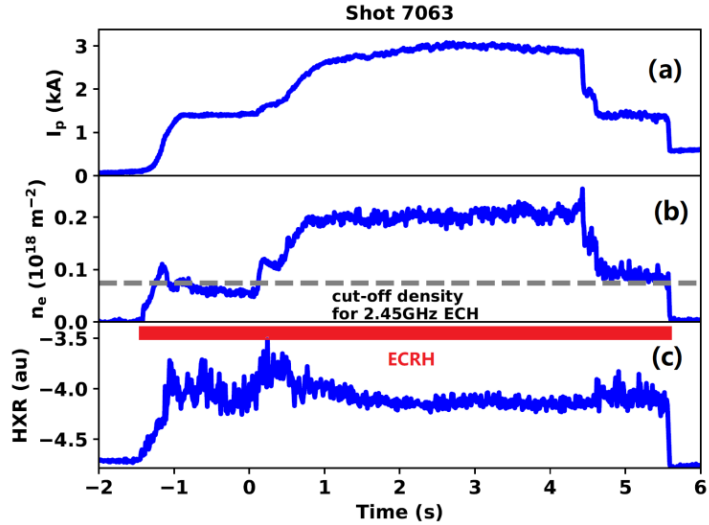

Fig.9 Waveforms of high density discharge with 2.45GHz ECRH. $P_{ECRH}$ was 20kW. The waveforms from top to bottom are （a）plasma current, Ip; (b) line integrated density; (c) intensity of hard x-ray.

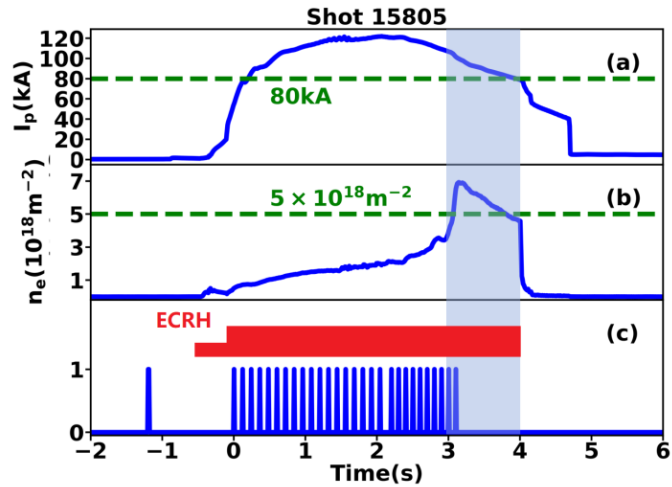

Fig.10 Waveforms of high density discharge with 28GHz ECRH. $P_{ECRH}$ was about 150kW. The waveforms from top to bottom are: (a) plasma current Ip; (b) line integrated density; (c) control signal of gas-puffing. One gyrotron injected 20kW from -0.5s to 1.5s and the other gyrotron injected 150kW from -1s to 4s.

## 6. Summary

New records of driven plasma current and current drive effectiveness have been obtained during the CS-free ECRH-only operation on EXL-50. The energetic electrons play a unique and important role in EXL-50's plasmas. Metal wall of the vacuum vessel effecting multiple reflections and absorption at high multi-harmonic resonances increase the high efficiency of acceleration of the energetic electrons. The asymmetric distribution of the energetic electrons in velocity space based on orbit analysis in a multi-fluid equilibrium is another key feature of the very high current drive effectiveness observed. However, the physics mechanism for the solenoid-free ECRH current drive is not yet well understood and quantifiable. Nevertheless, these results demonstrate an exceptional potential of ECRH to achieve and maintain highly efficient steady-state current drive over a density range up the ECRH cut-off density in the CS-free EXL-50 device. Theory, modelling and simulation to better describe the full range of mechanisms associated with the CS-free current drive observed will be systematically developed in the upcoming research. Steady-state high current and density experiments using high ECRH power on EXL-50 will make physics contributions of efficient current drive method for use in an eventual commercial ST fusion reactor.

**Acknowledgement**

**Appendix I. Simulation of hard x-ray spectrum based on three-temperature distribution model for energetic electrons**

The anisotropic velocity distribution function $f_s$ of the superthermal electrons in magnetic confined plasma can be represented by a three-temperature Maxwellian model (3T model) [32].

$$f_s(\vec{p}) = C_N exp\left(-\frac{p_\perp^2}{2T_\perp} - \frac{p_\parallel^2}{2T_{\parallel F}}\right) \text{ for } p_\parallel \geq 0$$

$$f_s(\vec{p}) = C_N exp\left(-\frac{p_\perp^2}{2T_\perp} - \frac{p_\parallel^2}{2T_{\parallel B}}\right) \text{ for } p_\parallel < 0$$

and $f_s(\vec{p}) = 0$, for $\vec{p} > p_c$

where $p_\parallel$ and $p_\perp$ are the parallel and perpendicular components of the momentum with respect to the magnetic field, normalized to the electron rest momentum $m_e c$; $T_\perp$, $T_{\parallel F}$ and $T_{\parallel B}$ are the perpendicular, parallel forward and parallel backward temperature, respectively, normalized to the electron rest mass $m_e c^2$; $p_c$ is the upper cut-off momentum ( $p_c = \sqrt{(10T_{\|F} + 1)^2 - 1}$ in simulation). For the runaway electrons, the perpendicular and parallel backward velocities are much smaller than the parallel forward velocity. So, $T_{\|F} \gg T_\perp$ and $T_{\|B}$ in 3T model can represent the distribution of runaway electrons. For the energetic electrons generated by RF heating, the perpendicular and parallel backward velocities are same order as the parallel forward velocity. $T_{\|F} \sim T_\perp$ and $T_{\|B}$ in 3T model is a good approximation for the distribution of energetic electrons in RF plasma.

Only the electron-ion bremsstrahlung radiation is considered for the hard x-ray simulation. The electron-electron bremsstrahlung and recombination radiation can be neglected compare to the electron-ion bremsstrahlung [32]. The x-ray is the line integrated measurement. The radial profile of ion density, energetic electron density and temperatures in 3T model are the key parameters to determine the shape of HX spectrum. For the simulation of HX spectrum in EXL-50, the energetic electron density is assumed as constant in thermal plasma region. The profiles of the ion density and energetic electron temperatures are assumed as parabolic distribution. $T_\perp / T_{\|F}$ and $T_{\|B} / T_{\|F}$ are 0.1 and 0.1 for runaway case, 1 and 0.75 and for ECRH case, respectively. The ratio of $T_\perp / T_{\|F}$ and $T_{\|B} / T_{\|F}$ is fixed and not change with radius. Under the above assumption, the best fitting of forward HX spectrum for 7841 in Fig.6 can be obtained when the peak value of $T_{\|F}$ is setting as 220keV for runaway case and 180keV for ECRH case.

## Appendix II. The EXL-50 Team

Minsheng Liu, Y-K Martin Peng, Baoshan Yuan, Yuejiang Shi, XianMing Song, Bing Liu, Shaodong Song, Xin Zhao, Enwu Yang, Wenwu Luo, Peihai Zhou, Yuanming Yang, Bo Xin, Yunyang Song, Dong Guo, Jiaqi Dong, Huasheng Xie, Yubao Zhu, Wenjun Liu, Xiang Gu, Di Luo, Bin Chen, Tiantian Sun, Zhi Li, Mingyuan Wang, Hanyue Zhao, Yukun Bai, Haojie Ma, Akio Ishida, Takashi Maekawa, Xiaorang Tian, Chao Wu, Guosong Zhang, Shunqiao Huang, Hao Li, Zimo Gao, Jiangbo Ding, Qing Zhou, He Liu, Lei Han, Pengmin Li, Hanqing Wang, Zhenxing Wang, Yupeng Guan, Zhen Li, Zihua Kang, Yong Liu, Xiuchun Lun, Dongkai Qi, Wei Wang, Quan Wu, Chunqi Liu, Kun Han, Yu Wang, Bo Chen, Renhua Bai, Xiaokun Bo, Pengmei Jia, Hong Zang, Yunfeng Gu, Jianbo Hao, Hefei Yuan, Xu Yang, Xiang Gao, Lianxing Chen, Shaoxun

Liu, Yumin Wang, Yingying Li, Shikui Cheng, Xianli Huang, Songjian Li, Xiaomin Tian, Renyi Tao, Hongyue Li, Bihe Deng, Qifeng Xie, Jiahe Liu, Chunrong Feng, Lingling Dai, Weiqiang Tan, Lin Chen, Ji Qi, Lihua Chen, Wenhao Cui, Shuo Wang, Song Liu, Haiwei Zhao, Afeng Zhao, Wei Yuan, Haiwei Zhao, Cece Dong, Fei Liu, Jun Zhao, Wenyin Li, Zequn Sun, Jianbo Hou, Huiling Wang, Chen Bian, Qinhai Wang, Xinxin Ma, Weigang Li, Jinmeng Dong, Jun Tan, Kaiming Feng, Houyang Guo, Zhenqi Zhu

**Reference**

[1] Artsimovich, L. A., Brobrovskii, G. A., Gorbunov, E. P., Ivanov, D. P., Kirillov, V. D., Kuznetsov, E. I., Mirnov, S. V., Petrov, M. P., Rasumova, K. A., Strelkov, V. S., and Shcheglov, D. A., Proc. Third Intern. Conf. on Plasma Physics and Nuclear Fusion, Novosibirsk, 1968, 1, 157 (Paper CN24/B1) (International Atomic Energy Agency, Vienna, 1969), Artsimovich, L. A. et al., Thermal insulation of plasma in the Tokamaks, 1967 Sov. At. Energy (Soviet Atomic Energy) 22 325-331

[2] N. J. Peacock, D. C. Robinson, M. J. Forrest, P. D. Wilcock & V. V. Sannikov, Measurement of the Electron Temperature by Thomson Scattering in Tokamak T3, Nature 224, 488 - 490 (01 November 1969)

[3] Peng Y.-K.M. and Strickler D.J., Features of spherical torus plasmas, Nucl. Fusion 26 769-777 (1986)

[4] Peng Y.-K.M., The physics of spherical torus plasmas, Phys. Plasmas 7, 1681-1692 (2000).

[5] Sykes A., Akers R., Appel L., Carolan P.G., Conway N.J., et al, High- performance of the START spherical tokamak , Plasma Phys. Control. Fusion 39, B247-260(1997)

[6] Synakowski E.J., Bell M. G., Bell R.E., Bigelow T., Bitter M., et al., The national spherical torus experiment (NSTX) research programme and progress towards high beta, long pulse operating scenarios, Nucl. Fusion 43 1653-1664 (2003)

[7] Buttery R. J., Akers R., Arends E., Conway N. J., Counsell G. F., et al., Stability at high performance in the MAST spherical tokamak, Nucl. Fusion 44 1027-1035 (2004)

[8] V.K. Gusev, S.E. Aleksandrov, V. Kh Alimov, I.I. Arkhipov, B.B. Ayushin, et al., Overview of results obtained at th Globus-M spherical tokamak, Nucl. Fusion 49 104021 (2009)

[9] Y.-K. M. Peng et al, P.R.China Patent pending (application no. 202010292584.X).

[10] C. B. Forest, Y. S. Hwang, M. Ono, and D. S. Darrow，Internally Generated Currents in a Small-Aspect-Ratio Tokamak Geometry, Phys. Rev. Lett. 68, 3559-3562 (1992)

[11] T. Maekawa, Y. Terumichi, H. Tanaka, M. Uchida, T. Yoshinaga, Formation of spherical tokamak equilibria by ECH in the LATE device, Nucl. Fusion 45 1439-1445 (2005)

[12] T. Yoshinaga, M. Uchida, H. Tanaka, and T. Maekawa, Spontaneous Formation of Closed-Field Torus Equilibrium via Current Jump Observed in an Electron-Cyclotron-Heated Plasma, Phys. Rev. Lett. 96, 125005 (2006)

[13] M. Uchida, T. Yoshinaga, H. Tanaka, T. Maekawa, Rapid Current Ramp-Up by Cyclotron-Driving Electrons beyond Runaway Velocity, Phys. Rev. Lett. 104, 065001 (2010)

[14] K. Kuroda, M. Wada, M. Uchida, H. Tanaka, T. Maekawa, Shift in principal equilibrium current from a vertical to a toroidal one towards the initiation of a closed flux surface in ECR plasmas in the LATE device, Plasma Phys. Control. Fusion 58, 025013 (2016)

[15] H. Tanaka, Y. Nozawa, M. Uchida, R. Kajita, Y. Omura, et al., Electron Bernstein wave heating and current drive with multi-electron cyclotron resonances during non-inductive start-up on LATE, in Proceedings of 27th IAEA Fusion Energy Conference, Ahmedabad, India (2018), p. EX/P3-19

[16] A. Ejiri, Y. Takase, T. Oosako, T. Yamaguchi, Y. Adachi, et al., Non-inductive plasma current start-up by EC and RF power in the TST-2 spherical tokamak, Nucl. Fusion 49 065010 (2009)

[17] Y.Takase, A. Ejiri, H. Kakuda, T. Oosako, T. Shinya,, Non-inductive plasma initiation and plasma current ramp-up on the TST-2 spherical tokamak, Nucl. Fusion 53 063006 (2013)

[18] V.F. Shevchenko, M.R. O'Brien, D. Taylor, A.N. Saveliev, Electron Bernstein wave assisted plasma current start-up in MAST, Nucl. Fusion 50 022004 (2010)

[19] V. F. Shevchenko, T. Bigelow, J. B. Caughman, S. Diem, J. Mailloux, et al., Long Pulse EBW start-up experiments in MAST, EPJ Web Conf. 87, 02007 (2015)

[20] K. Hanada, H. Zushi, H. Idei, K. Nakamura, M. Ishiguro, et al., Non-Inductive Start up of QUEST Plasma by RF Power, Plasma Science and Technology, 13, 307-311 (2011)

[21] M. Ishiguro, K. Hanada, H. Liu, H. Zushi, K. Nakamura, et al., Non-inductive current start-up assisted by energetic electrons in Q-shu University experiment with steady-state spherical tokamak, Phys. Plasmas 19, 062508 (2012)

[22] S. Tashima, H. Zushi, M. Isobe, K. Hanada, H. Idei, et al., Role of energetic electrons during current ramp-up and production of high poloidal beta plasma in non-inductive current drive on QUEST, Nucl. Fusion 54, 023010 (2014)

[23] H. Idei, T. Kariya, T. Imai, K. Mishra, T. Onchi, et al., Fully non-inductive second harmonic electron cyclotron plasma ramp-up in the QUEST spherical tokamak, Nucl. Fusion 57,126045 (2017).

[24] H. Idei, T. Onchi, T. Kariya, et al., fully non-inductive 2nd harmonic electron cyclotron current ramp-up with polarized focusing-beam in the quest spherical tokamak, in Proceedings of 27th IAEA Fusion Energy Conference, Ahmedabad, India (2018), EX/P3-21

[25] H. Idei, T. Onchi, K. Mishra, H. Zushi, T. Kariya, et al., Electron heating of over-dense plasma with dual-frequency electron cyclotron waves in fully non-inductive plasma ramp-up on the QUEST spherical tokamak, Nucl. Fusion 60, 016030 (2020)

[26] T. Onchi, H. Idei, M. Fukuyama, D. Ogata, R. Ashida, et al., Non-inductive plasma current ramp-up through oblique injection of harmonic electron cyclotron waves on the QUEST spherical tokamak, Phys. Plasmas 28, 022505 (2021)

[27] S.J. Li, R.H. Bai, R.Y. Tao, N. Li, X.C. Lun, L.C. Liu, Y. Liu, M.S. Liu and B.H. Deng, A quasi-optical microwave interferometer for the XuanLong-50 experiment, Journal of Instrum. 16, T08011(2021)

[28] S. K. Cheng, Y. B. Zhu, Z. Y. Chen, Y. X. Li, R. H. Bai, B. Chen, X. L. Huang, L. L. Dai,and M. S. Liu, Tangential hard x-ray diagnostic array on the EXL-50 spherical tokamak, Rev. Sci. Instrum. 92, 043513 (2021)

[29] C. Angioni, E. Fable, M. Greenwald, M. Maslov, A. G. Peeters, H. Takenaga and H. Weisen, Particle transport in tokamak plasmas, theory and experiment, Plasma Phys. Control. Fusion 51, 124017 (2009)

[30] N.J.Fisch, and A.H. Boozer, Creating an Asymmetric Plasma Resistivity with Waves, Phys. Rev. Lett. 45, 720-722 (1980）

[31] T. Ohkawa, Steady-state operation of tokamaks by r-f heating, General Atomics Report No. GA-A13847 (1976).

[32] J.Stevens, S. Von Goeler, S. Bernabei, M. Bitter, T.K.Chu, et al., Modelling of the electron distribution based on bremsstrahlung emission during lower-hybrid current drive on PLT, Nucl. Fusion 29, 1529-1541 (1985)

[33] A. Ishida, Y.K. M. Peng, W. J. Liu, Four-Fluid Axisymmetric Plasma Equilibrium Model Including Relativistic Electrons and Computational Method and Results, Phys. Plasmas 28, 032503 (2021)

[34] T. Maekawa, T. Yoshinaga, M. Uchida, F. Watanab and H. Tanaka，Open field equilibrium current and cross-field passing electrons as an initiator of a closed flux surface in EC-heated toroidal plasmas，Nucl. Fusion 52, 083008 (2012)

[35] D.Guo, Y.J.Shi, et al., Experimental study on edge energetic electrons in EXL-50 spherical torus, https://arxiv.org/abs/2112.10315